\documentclass[sigconf,nonacm]{acmart}
\usepackage{algorithmic,algorithm}
\usepackage{booktabs}
\usepackage{multirow}
\usepackage{graphicx}
\usepackage{textcomp}
\usepackage{xcolor}
\usepackage{enumitem}

\newcommand{\bh}[1]{\vspace{2pt}\noindent \textbf{#1} }
\newcommand{\ih}[1]{\vspace{2pt}\noindent \textit{#1} }

\AtBeginDocument{%
  }

\newboolean{comment_bullets}
\setboolean{comment_bullets}{False}

\begin{document}



\title{DUET: Disaggregated Hybrid Mamba-Transformer LLMs with Prefill and Decode-Specific Packages}


\author{Alish Kanani\textsuperscript{1},
        Sangwan Lee\textsuperscript{2},
        Han Lyu\textsuperscript{1},
        Jiahao Lin\textsuperscript{1},
        Jaehyun Park\textsuperscript{2},
        Umit Y.\ Ogras\textsuperscript{1}}
\affiliation{%
  \institution{\textsuperscript{1}University of Wisconsin--Madison \quad \textsuperscript{2}University of Ulsan}
  \city{}
  \state{}
  \country{}
}
\email{
{ahkanani, hlyu42, jlin445, uogras}@wisc.edu, {lso500,jaehyun}@ulsan.ac.kr}

\begin{abstract}
Large language models operate in distinct \textit{compute-bound prefill} followed by memory \textit{bandwidth-bound decode} phases.
Hybrid Mamba– Transformer models inherit this asymmetry while adding state-space model (SSM) recurrences and element-wise operations that map poorly to matmul-centric accelerators.
This mismatch causes performance bottlenecks, showing that \textit{a homogeneous architecture cannot satisfy all requirements}.
We introduce DUET, a disaggregated accelerator that assigns prefill and decode phases to specialized packages. 
The Prefill package utilizes systolic array chiplets with off-package memory for efficient large matrix multiplications and long-sequence SSMs.
The Decode package utilizes vector-unit arrays with high-bandwidth in-package memory to accelerate token-by-token SSM and vector–matrix multiplications.
Both architectures are runtime-configurable to support hybrid models with mixed Mamba and attention layers.
Evaluations on Nemotron-H-56B, Zamba2-7B, and Llama3-8B across four workloads show that DUET achieves 4$\times$ faster time to first token, 1.4$\times$ higher throughput, and 1.5$\times$ lower time between tokens over the B200 GPU. 
\end{abstract}

\maketitle



\vspace{-2mm}
\section{Introduction}\label{sec:intro}


Auto-regressive large language models (LLMs) have become foundational across applications, yet their inference pipelines still suffer from a fundamental asymmetry. 
The prefill (prompt processing) phase is compute-bound and highly parallel, whereas the decode (token generation) phase is memory bandwidth-bound and inherently sequential due to intensive key-value (KV) cache accesses. 
Modern LLM serving frameworks observe this asymmetry and often assign prefill and decode to different GPUs~\cite{zhong2024distserve,qin2025mooncake}. 
However, these frameworks remain confined to homogeneous GPUs, which cannot fully exploit the distinct phase characteristics.
As a result, the prefill phase continues to underutilize memory bandwidth, while the decode phase underutilizes compute units.
This persistent mismatch motivates a \emph{hardware-level disaggregation}, where each phase is executed on a package specialized for its dominant bottleneck.

This challenge persists in emerging hybrid Mamba–Transformer architectures.
Hybrid models combine the complementary strengths of state-space models (SSMs) and self-attention~\cite{blakeman2025nemotron, lieber2024jamba, glorioso2024zamba, ibm_bamba_2025}. 
SSMs offer efficient linear-time computation with a fixed-size state representation, whereas attention excels at long-range context retrieval. 
Nevertheless, they remain autoregressive during inference and retain the same compute-bound prefill followed by a sequential, memory bandwidth-bound decode phase.

\begin{figure}[t]
\centering
\includegraphics[width=1\linewidth]{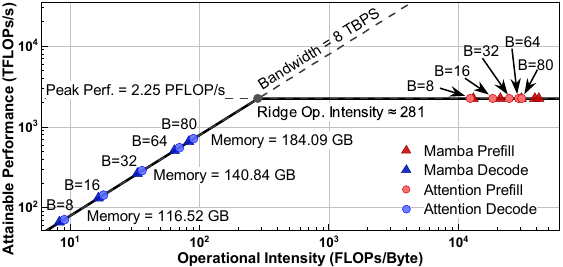} 
\vspace{-7mm}
\caption{Roofline analysis of Mamba and Transformer layers from Nemotron-H-56B~\cite{blakeman2025nemotron} as function of batch size (B) on Nvidia Blackwell B200
(2.25 PFLOP/s peak, 8 TB/s HBM bandwidth, 192 GB capacity)~\cite{nvidia_blackwell_datasheet_2024}.}
\vspace{-10mm}
\label{fig:roofline}
\end{figure}

Our roofline analysis of Nemotron-H-56B~\cite{blakeman2025nemotron} quantifies this asymmetry across both Mamba and attention layers.
We model per-layer compute and memory traffic at a 4K context length and project the resulting operational intensities (FLOPs per byte) onto the FP16 roofline of an NVIDIA Blackwell B200 GPU~\cite{nvidia_blackwell_datasheet_2024}. 
Despite using different kernels, Mamba and attention layers exhibit similar operational intensity across the prefill and decode phases, as shown in Figure~\ref{fig:roofline}. 
Prefill falls in the compute-bound region, while the decode remains bandwidth-bound, even with a batch size of 80 (which already consumes $\sim$184 GB). 
These results lead to two key insights: 
(1) prefill and decode require phase-specialized hardware; and 
(2) each phase must efficiently support both SSM and attention kernels, since hybrid models interleave them. 
These results, combined with the inefficiency of mapping SSM recurrences on matmul-centric GPUs, highlight \textit{the need for disaggregated hardware that jointly addresses compute, memory, and kernel diversity}.

To address the bottlenecks of the prefill and decode phases in hybrid models simultaneously, we propose \emph{DUET, a disaggregated acceleration framework that tailors the hardware for the specific needs of prefill and decode operations.}
For the compute-bound prefill phase, it employs systolic-array–based chiplets with off-package memory to execute large matrix–matrix (GEMM) operations.  
For the memory bandwidth-bound decode phase dominated by matrix-vector (GEMV) operations, it employs vector-unit array-based chiplets coupled with in-package high bandwidth memory (HBM).
While systolic arrays and vector units are well-suited for GEMM and GEMV multiplications, respectively, SSMs introduce recurrent dependencies and element-wise operations that challenge their direct mapping. 
DUET addresses this challenge with novel microarchitectural extensions that enable systolic arrays to process long-sequence SSMs efficiently while preserving their efficiency on standard GEMM multiplications. It also extends vector units to support token-by-token SSM execution without reducing their efficiency on GEMV multiplications. 
Since hybrid LLMs interleave SSM and attention blocks, the proposed systolic and vector-unit array architectures are runtime-configurable. 
In summary, \textit{the DUET framework advances both system-level (disaggregated prefill–decode acceleration) and circuit-level (configurable microarchitectures) design methodology for efficient inference of emerging hybrid LLMs.}

The DUET framework supports a wide range of hardware configurations for both Prefill and Decode packages. We first perform a comprehensive design space exploration for both systolic and vector-unit arrays to identify the Pareto-optimal configurations. Then, we construct a concrete DUET system that co-optimizes the performance-area trade-off.
We evaluate this DUET system on two hybrid models (Nemotron-H-56B~\cite{blakeman2025nemotron} and Zamba2-7B~\cite{glorioso2024zamba2}) and a pure Transformer model (Llama3-8B~\cite{grattafiori2024llama3}) on ArXiv, BWB, LongWriter, and Chat inference workloads.
DUET is compared against NVIDIA’s B200~\cite{nvidia_blackwell_datasheet_2024} as well as two aggregated baselines: a prefill-friendly system with off-package memory and a decode-friendly system with in-package HBM.
Across all workloads, DUET consistently outperforms all baselines, reducing Time-to-First-Token (TTFT) by 4$\times$, 1.4$\times$, and 2.7$\times$ during the prefill and improving decode throughput by 1.4$\times$, 3.3$\times$, and 1.1$\times$ while reducing Time-Between-Token (TBT) by 1.5$\times$, 4$\times$, and 1.2$\times$.
These results highlight the practicality and efficiency of the proposed disaggregated architecture.
\textit{The key contributions of this work are as follows:}
\begin {itemize}[topsep=2pt, leftmargin=*]
\item A disaggregated acceleration framework for hybrid LLMs, with compute and memory tailored for prefill and decode phases.
\item Runtime configurable microarchitectures, including systolic arrays for long GEMM/SSM kernels and vector-unit arrays for token-wise GEMV multiplication and SSM kernels.
\item Comprehensive evaluations across hybrid and Transformer models on diverse workloads, demonstrating consistent improvements in TTFT, throughput, and TBT.
\end {itemize}

The cycle-accurate simulator for configurable systolic and vector unit arrays is available at \url{https://github.com/AlishKanani/DUET}.

\vspace{-5mm}
\section{Related Work}\label{sec:related}
Disaggregated inference frameworks, such as DistServe~\cite{zhong2024distserve} and Mooncake~\cite{qin2025mooncake}, separate the two phases across different GPUs. 
While this approach improves TTFT for prefill and throughput for decode, \textit{both phases still rely on identical GPUs}, 
leading to underutilized memory bandwidth during prefill and idle compute units during decode. 
This inefficiency motivates hardware-level disaggregation, where each phase is mapped to a GPU optimized for its dominant bottleneck. 
SPAD introduces distinct prefill and decode chips, reporting cost advantages over GPUs~\cite{zhang2025spad}. 
However, it targets only Transformer workloads and retains a monolithic design, which is not scalable, as summarized in Table~\ref{tab:related}.

Hybrid Mamba–Transformer models outperform Transformer baselines while keeping similar accuracy.
For example, Nemotron-H (NVIDIA) achieves up to 3$\times$ faster inference at comparable quality~\cite{blakeman2025nemotron}, Jamba (AI21) reports an 8–10$\times$ reduction in KV-cache footprint~\cite{lieber2024jamba}, and Bamba (IBM) demonstrates 2$\times$ speedup over similar-sized Transformers~\cite{ibm_bamba_2025}. 
These models typically employ a 3:1–7:1 ratio of Mamba to attention blocks, reflecting a growing shift toward Mamba-dominant hybrid LLMs.

Recent efforts to accelerate Mamba models follow two main directions. 
Unified matrix multiplication and SSM units, such as MARCA~\cite{li2024marca,li2025marca}, HCSAs~\cite{jin2025hcsas}, HLX~\cite{jung2025hlx}, and EpochCore~\cite{raja2025systolic}, retain matrix multiplication compatibility, as shown in Table~\ref{tab:related}.
Their implementation involves frequent reads/writes for intermediate element-wise operations, achieving limited gains. 
Conversely, dedicated SSM accelerators, including LightMamba~\cite{wei2025lightmamba}, FastMamba~\cite{wang2025fastmamba}, SSM-RDU~\cite{ko2025ssm}, eMamba~\cite{kim2025emamba}, and SpecMamba~\cite{zhong2025specmamba}, design specialized processing units for state updates. 
They achieve high efficiency for pure Mamba workloads, but the same compute unit does not support matrix multiplication. 
Notably, none of these works address the compute/memory asymmetry between the prefill and decode phases, as well as hybrid execution patterns together.


In contrast to prior work, DUET combines phase-specific hardware with unified SSM-attention execution. 
The Prefill and Decode packages eliminate redundant read/write of intermediate element-wise SSM updates via in-situ fusion. 
Built as a chiplet-based framework, DUET also provides modularity and scalability, offering a robust path forward for emerging hybrid LLMs.

\begin{table}[t]
\caption{Comparison of state-of-the-art LLM accelerators.}
\vspace{-4mm}
\label{tab:related}
\resizebox{\linewidth}{!}{%
\begin{tabular}{c|c|c|c|c}
\hline
 & \textbf{\begin{tabular}[c]{@{}c@{}}Disagg-\\ 
 regated
 \end{tabular}} & \textbf{\begin{tabular}[c]{@{}c@{}}\hspace{-1mm}Unified SSM\hspace{-1mm} \\ \& matmul\end{tabular}} & \textbf{\begin{tabular}[c]{@{}c@{}}Chiplet \\ based\end{tabular}} & \textbf{\begin{tabular}[c]{@{}c@{}} Evaluated\\ Model\end{tabular}} \\ \hline
SPAD~\cite{zhang2025spad} & $\checkmark$ & $\times$ & $\times$ & Transformer only \\ \hline
MARCA~\cite{li2024marca} & $\times$ & $\checkmark$ & $\times$ & Mamba only \\ \hline
HLX~\cite{jung2025hlx} & $\times$ & $\checkmark$ & $\times$ & Hybrid (Transformer + Mamba) \\ \hline
\hspace{-1mm}EpochCore~\cite{raja2025systolic}\hspace{-1mm} & $\times$ & $\checkmark$ & $\times$ & S4, Liquid-S4 \\ \hline
\textbf{DUET} & $\pmb\checkmark$ & $\pmb\checkmark$ & $\pmb\checkmark$ & \hspace{-1mm}\textbf{Hybrid  (Transformer + Mamba})\hspace{-1mm} \\ \hline
\end{tabular}
}
\vspace{-5mm}
\end{table}
\vspace{-2mm}
\section{Disaggregated DUET Architecture}\label{sec:duet}
\vspace{-0.5mm}
\subsection{System Overview}\label{ssec:overview}
\vspace{-0.5mm}

\begin{figure*}[t]
\centering
    \centering
    \includegraphics[width=\linewidth]{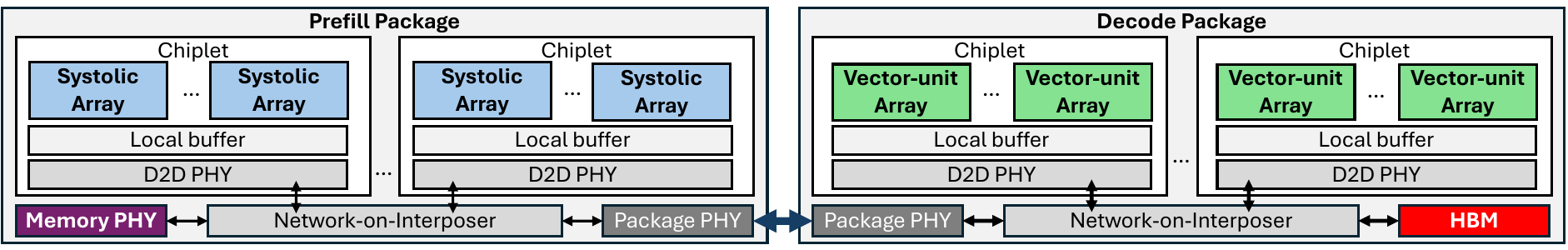}
    \vspace{-7mm}
    \caption{Overview of the DUET disaggregated acceleration framework. The Prefill package uses configurable systolic-array chiplets with off-package DRAM, while the Decode package employs configurable vector-unit chiplets with in-package HBM.}
    \vspace{-3mm}
    \label{fig:overview}
\end{figure*}

The proposed disaggregated acceleration framework consists of two heterogeneous packages: a Prefill package optimized for compute-bound execution and a Decode package optimized for memory bandwidth-bound execution, as shown in Figure~\ref{fig:overview}. 
Both packages are realized as chiplet-based systems, enabling phase-specific specialization while staying within the practical limits of advanced packaging.
We adopt chiplets for two reasons:
(1) advanced-node reticle limits and yield constraints make assembling multiple small dies more practical than fabricating a single large monolithic chip~\cite{stow2016cost,jaiswal2024halo,ayari2016schedulability}; and 
(2) chiplets allow independent specialization of compute and memory subsystems, 
on a common interposer~\cite{kanani2025thermos,qi2023moela,sharma2025hemu,taheri2024red}. 

\bh{Prefill Package:} 
Since the Prefill package targets the compute-bound phase, it prioritizes silicon area for dense systolic computation. 
As illustrated in Figure~\ref{fig:overview}, it integrates multiple compute chiplets, each containing systolic arrays, local buffers, and die-to-die (D2D) PHYs connected through a moderate-bandwidth network-on-interposer (NoI). 
Surrounding memory-PHY chiplets interface with off-package GDDR stacks, providing sufficient bandwidth to sustain prefill workloads.
This configuration maximizes compute density while keeping memory interfaces compact, as the prefill phase imposes relatively modest bandwidth demands.

\bh{Decode Package:}
Since the Decode package is designed for the bandwidth-bound phase, it emphasizes proximity between compute and memory to minimize data movement.
As shown in Figure~\ref{fig:overview}, it integrates multiple vector-unit chiplets interleaved with HBM stacks on a shared interposer, enabling short, high-bandwidth connections through a dense NoI.
It also allocates a larger portion of its silicon and interposer resources to memory interfaces and D2D PHYs than the Prefill package due to higher bandwidth demands.
This tightly coupled layout provides the bandwidth required to sustain SSM state and KV-cache accesses during the decode phase.

\bh{Inter-Package Connectivity and Scalability:}
The Prefill and Decode packages communicate activations (the embedding output generated by the prefill stage) and caches (KV- and SSM state cache) via package-level interconnects such as NVLink~\cite{wei20239} and UALink~\cite{ualink2025_200g}.
Although off-package links offer lower bandwidth than an on-interposer network, their latency impact is effectively hidden.
The cache transfers can be overlapped with computations in the next layer because LLM inference progresses layer-by-layer. 
With parameterized chiplet counts and systolic/vector-unit array sizes, the DUET framework enables scalable architectures.

\vspace{-3mm}
\subsection{Prefill Package: Configurable Systolic Array}\label{ssec:prefill}


\bh{Goals and design requirements:}
The compute-oriented Prefill package requires:
(1) a unified compute array that supports both SSM and GEMM execution;
(2) a dataflow that keeps the recurrent state local and avoids external SRAM traffic for SSM intermediates; and
(3) a lightweight control mechanism that reconfigures the array between SSM and GEMM dataflows.

\bh{Systolic array microarchitecture:}
Each compute chiplet integrates multiple systolic arrays, 
where each processing element (PE) in an array contains an MAC and a small set of local registers.
Compared to conventional systolic design with two registers~\cite{kung1979systolic,genc2021gemmini}, we use two additional registers to store SSM-specific intermediate values and a lightweight control logic. 
Neighbor links enable one-word-per-cycle propagation in both the horizontal (W$\rightarrow$E) and vertical (N$\rightarrow$S) directions. 
A single column of SFUs placed alongside the array provides configurable non-linear operations. 
These SFUs read operands from on-chip SRAM buffers and feed the results directly into the PE pipeline, allowing intermediate values to be consumed in situ without SRAM write-backs.

\begin{table}[b]
\centering
\vspace{-5mm}
\caption{Key variables used in Mamba-SSM models.}
\label{tab:notation}
\vspace{-3mm}
\resizebox{\linewidth}{!}{%
\begin{tabular}{c|l}
\hline
\textbf{Notation} & \textbf{Definition} \\
\hline
$ u_k, y_k $ & Input, output embedding for token $k$ \\
$ \Delta_k $ & Input-dependent discretization step \\
$ A $ & Continuous-time state transition Matrix \\
$ \mathbf{B}_k,\mathbf{C}_k $ & Input-dependent input-to-state, state-to-output matrix \\
$ D $ & Direct feedthrough term \\
$ \mathbf{X}_k $ & Recurrent state \\
$ \bar{A}_k = \exp(\Delta_k A) $ & Discretized state transition matrix \\
$\bar{u}_k = \Delta_k u_k$ & Discretized input embedding \\
ED & Embedding dimension \\
N & State dimension \\
\hline
\end{tabular}
}
\end{table}

\begin{figure}[t]
\centering
\includegraphics[width=\linewidth]{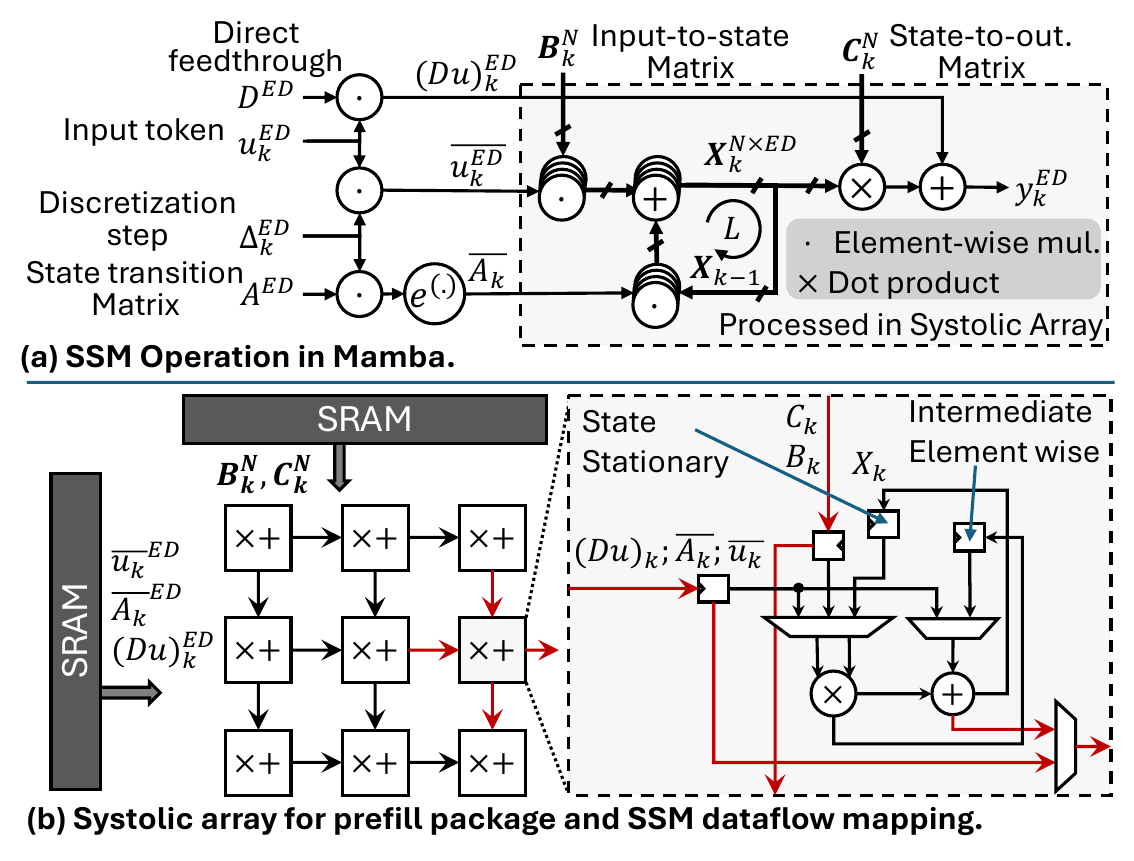}
\vspace{-8mm}
\caption{Overview of SSM computation~\cite{dao2024transformers} and its state-stationary dataflow mapping on systolic array. This operation repeats over the sequence length $L$.} 
\label{fig:sa_dataflow}
\vspace{-8mm}
\end{figure}

\bh{State-stationary SSM-prefill mode:}
SSMs fundamentally differ from GEMM-based operations because they involve element-wise recurrences with strict inter-step dependencies~\cite{gu2024mamba,dao2024transformers}.
As shown in Figure~\ref{fig:sa_dataflow}(a), for each token, the SSM combines the previous state with the current input, producing a new state $\mathbf{X}_k$ and output $y_k$:
\vspace{-1mm}
\begin{equation}\label{eq:ssm}
    \mathbf{X_k} =  \exp(\Delta_k A) \mathbf{X}_{k-1} + (\Delta_k \cdot \mathbf{B}_k)u_k;\quad
     y_k = \mathbf{C}_k \times \mathbf{X_k} + D \cdot u_k
\vspace{-1mm}
\end{equation}
where all variables are defined in Table~\ref{tab:notation}. The input-to-state matrix $\mathbf{B}_k$, state-to-output matrix $\mathbf{C}_k$, and discretization step $\Delta_k$ 
are input-dependent parameters generated via linear projections of the input embedding $u_k$. 
As shown in Figure~\ref{fig:sa_dataflow}(a), we further adopt a simple but effective algebraic reordering: prior Mamba formulations~\cite{gu2024mamba,dao2024transformers} express the term as $(\Delta_k \cdot \mathbf{B}_k)u_k$, which we reorder as $(\Delta_k \cdot u_k)\mathbf{B}_k$.
Since scalar–vector multiplication is associative, this reordering preserves correctness while reducing two vector-wide multiplications to one vector-wide and one scalar multiplication.

SSM operates along two axes: the embedding dimension (ED) and the state dimension (N). As illustrated in Figure~\ref{fig:sa_dataflow}(b), we unroll ED across the array rows and N across the columns. 
SSM parameters exhibit useful sharing properties.
The discrete state transition matrix $\bar{A}_k$, discrete input $\bar{u}_k$, and partial sums $(Du)_k$, are shared across columns, while $\mathbf{B}_k$ and $\mathbf{C}_k$ are shared across rows.
Critically, the recurrent state $\mathbf{X}_k$ is generated during prefill rather than fetched from memory, enabling a \emph{state-stationary dataflow} where each PE holds one state element locally.
During steady-state, each PE: 
\begin{enumerate}[nosep,leftmargin=*]
    \item receives $\mathbf{B}_k$, $\mathbf{C}_k$ vertically and $\bar{A}_k$, $\bar{u}_k$, and $(Du)_k$ horizontally;
    \item performs element-wise multiplication for $B_k\bar{u}_k$, $\bar{A}_kX_k$, and partial product $C_k X_k$ in a three-cycle micro-pipeline; and
    \item forwards partial sums horizontally to accumulate the output $y_k$ while the updated state remains stationary.
\end{enumerate}

In our architecture, all element-wise intermediates are consumed in-place, eliminating external SRAM traffic. 
Only input parameters and activations are streamed into the array.
After an initial $\mathcal{O}(N)$ pipeline fill, the array sustains one SSM update every three cycles. 
By keeping the recurrent state in the array and fusing all element-wise operations into the PE pipeline, the state-stationary dataflow maintains high utilization during long-sequence prefill while avoiding the typical read-write bottlenecks during SSM acceleration.





\noindent\textbf{Matrix-multiplication mode:}
The same systolic array executes GEMM for attention and feed-forward layers using an output-stationary dataflow. 
Inputs stream from the array edges, and partial sums accumulate within the PEs. The two SSM-specific registers remain clock-gated during GEMM execution. 
Hence, the GEMM's performance matches a conventional systolic array design. 
Switching between SSM and GEMMs is controlled via mode bits in the array controller; operand tiling, buffering, and addressing remain unchanged, thereby avoiding the need for expensive re-layout.


\vspace{-1mm}
\subsection{Decode Package: Configurable Vector-Unit}\label{ssec:decode}
\vspace{-1mm}

\bh{Goals and design requirements:}
The memory bandwidth-oriented Decode package requires: 
(1) high sustained memory bandwidth and short paths from memory to compute, 
(2) fast vector reductions (for $\mathbf{C}\mathbf{X}$ in SSM and GEMV multiplications), and 
(3) zero-overhead of element-wise SSM operations to avoid write-backs of intermediates. 
Systolic arrays are unsuitable here, as they incur $\mathcal{O}(N)$ pipeline latency and cannot quickly load the entire cache.

\begin{figure}[b]
\centering
\vspace{-5mm}
\includegraphics[width=0.9\linewidth]{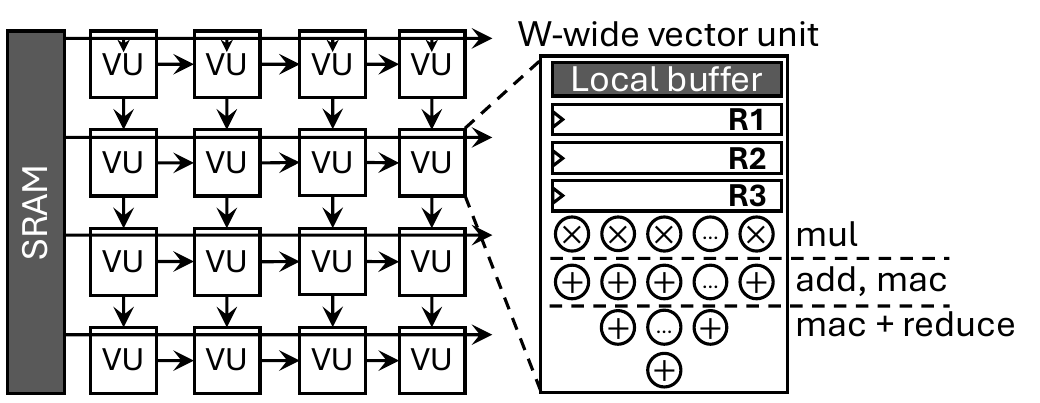}
\vspace{-4mm}
\caption{Vector-unit array for Decode package.}
\label{fig:va_dataflow}
\end{figure}

\bh{Vector-unit array microarchitecture:}
The core is an array of W-wide vector units. 
Each vector-unit supports vector–vector mul/add and a MAC + reduction datapath for dot products.
It includes a local buffer and three vector registers: two input registers and one register to hold intermediate element-wise SSM results, as shown in Figure~\ref{fig:va_dataflow}. 
A single SFU row adjacent to the array performs non-linear operations (e.g., \texttt{exp}, \texttt{SiLU}).

\bh{SSM-decode dataflow on the vector-unit array:}
The array consumes the state cache $\mathbf{X}_k$ (loaded into local buffer with a bypass path from external SRAM for fast initialization) together with the SSM parameters (defined in Table~\ref{tab:notation}) $\bar{A}_k, \bar{u}_k, \mathbf{B}_k, \mathbf{C}_k,$ and $D_k$.
For each token $k$, a vector-unit executes:
\begin{enumerate}[nosep,leftmargin=*]
\item vector-wise $\bar{A}_k\mathbf{X}_k$, $\mathbf{B}_k\bar{u}_k$, and update the state; and
\item compute $\mathbf{C}_k\times\mathbf{X}_k$ via the dot-product reduction path to produce output token $y_k$.
\end{enumerate}

The decode stage has only one SSM step per token, so the reduction latency is critical. 
The dot product latency is $\mathcal{O}(logN)$ due to its tree structure.
When N>W, multiple units (N/W) in a column process disjoint state slices and combine their partial results $(\mathbf{CX})$ via a simple scalar MAC chain (one MAC per vector-unit). 
This design keeps multipliers busy every cycle while avoiding read-writes of intermediate element-wise products back to SRAM.

\bh{Vector–matrix multiplication:}
The same vector units execute vector-matrix multiplications required in Transformer (e.g., token $\times$ FFN weights, query $\times$ key cache). 
The input vector is loaded once into the local buffer, and the matrix stream from external SRAM.
Outputs are calculated by the same dot-product/reduction path used for $\mathbf{CX}$ in SSM mode; therefore, no microarchitectural or data layout changes are required between SSM and GEMV modes.

Together, the vector-unit array and high-bandwidth memory form a decode engine well-suited for cache-heavy, token-wise SSM and GEMV workloads, all within a single efficient microarchitecture.
As a result, the Decode package complements the Prefill package and sustains high performance in the bandwidth-bound phase.

\vspace{-2mm}
\section{Experimental Evaluation}\label{sec:eval}

\subsection{Experimental Setup}\label{ssec:setup}

\bh{Evaluation Methodology:}
The FP16 systolic and vector-unit arrays are implemented using SystemVerilog and synthesized in Synopsys Design Compiler using a TSMC 28 nm standard-cell library, since 16-bit data format is the predominant choice in contemporary LLMs.
On-chip SRAM buffers are modeled using CACTI 7.0~\cite{balasubramonian2017cacti}.

\ih{Full-stack LLM simulation:}
LLM inference is evaluated using an in-house simulator that models cycle-accurate microarchitectures for the systolic and vector-unit arrays (validated against RTL).
Memory behavior and NoI communication are modeled using an event-driven simulation, which tracks per-chiplet queues and link utilization.
DRAM behavior is modeled using Ramulator-derived timing traces~\cite{luo2023ramulator}. 
All baseline systems are evaluated using the same simulation framework to ensure a fair comparison.

\bh{Models and Workloads:}
We evaluate three models: Nemotron-H-56B~\cite{blakeman2025nemotron}, Zamba2-7B~\cite{glorioso2024zamba2}, and Llama3-8B~\cite{grattafiori2024llama3} 
with four workloads: 
\begin{enumerate}[nosep,leftmargin=*]
    \item ArXiv-4K~\cite{cohan2018discourse} (ArXiv): long scientific documents, representing long prefill / short decode;
    \item Bilingual Web Books~\cite{jiang2023discourse} (BWB): multilingual narrative text, representing long prefill / long decode;
    \item LongWriter-6K~\cite{bailongwriter} (LongWriter): extended-context writing tasks, representing short prefill / long decode; and
    \item LMSYS-Chat-1M~\cite{zheng2023lmsys} (Chat): diverse chat interactions, representing short prefill / short decode.
\end{enumerate}

\vspace{-3mm}
\subsection{DSE for Systolic and Vector-unit Arrays}\label{ssec:dse}
The DUET framework enables scalable designs with different design trade-offs. 
We start with a design space exploration to determine the dimensions of the systolic and vector-unit arrays used in prefill and decode packages. 

\begin{figure}[t]
\centering
\includegraphics[width=\linewidth]{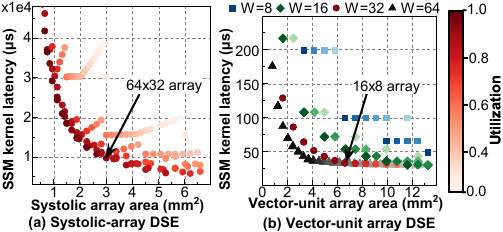}
\vspace{-7mm}
\caption{Design space exploration for a single SSM kernel in Nemotron-H-56B~\cite{blakeman2025nemotron}. (a) Systolic array used in the prefill phase, (b) Vector-unit array used in the decode phase.} 
\vspace{-4mm}
\label{fig:dse}
\end{figure}

\bh{Systolic Array:}
We sweep array dimensions by independently varying rows and columns across a range from $8\times8$ up to $256\times256$ under a fixed 256 GB/s SRAM bandwidth, reflecting that the SRAMs are backed by low-bandwidth GDDR.
Then, we evaluate performance for a long-context SSM kernel from Nemotron-H-56B with a sequence length of 2048.
While smaller arrays such as $8\times8$ minimize area, their latency is prohibitively high.
Conversely, very large arrays suffer from reduced utilization due to bandwidth bottlenecks.
Other SSM kernel sizes show similar behavior, indicating that these observations are not workload-specific.
From the resulting Pareto frontier in Figure~\ref{fig:dse}(a), we select a $64\times32$ systolic array, which offers low latency and high utilization.
We also observe the same trend for GEMM kernels and omit those results for brevity.


\bh{Vector-unit Array:}
We perform single-token execution and sweep both the array dimensions and vector-unit widths under a 1,024 GB/s SRAM bandwidth, reflecting that these SRAMs are fed by high-bandwidth HBM.
We vary $W \in \{8, 16, 32, 64 \}$ and explore array sizes ranging from $4\times×4$ to $32\times32$.
Smaller vector-unit widths (e.g., $W=8$ or 16) reduce per-unit parallelism and require more vector units to collaborate to form full vector operations, thus degrading latency.
Conversely, $W=64$ leads to underutilization for SSM variants whose state dimensions fall below 64.
From the Pareto frontier in Figure~\ref{fig:dse}(b), 
we select a $16\times8$ array with $W=32$, which provides a strong balance of utilization and latency.
Vector–matrix kernels exhibit similar behavior, indicating that this configuration is robust across workloads.
Power and achievable operating frequency are analyzed in Section~\ref{ssec:area_power}.

\vspace{-2mm}
\subsection{Optimized DUET System and Baselines}
Both Prefill and Decode packages are implemented as chiplet-based systems and fit within the practical maximum interposer area of roughly 2700–3000 $mm^2$ \cite{lau2022recent}.
This is comparable to state-of-the-art accelerator packages such as NVIDIA B200~\cite{nvidia_blackwell_datasheet_2024} and AMD MI350 \cite{amd_mi350_series_2025}.
We limit all chiplets to 121 $mm^2$, matching the HBM footprint to simplify floorplanning~\cite{park2022192}.

\bh{Prefill package:}
The Prefill package integrates 16 compute chiplets arranged in a $4\times4$ grid as summarized in Table~\ref{tab:spec}. 
Each chiplet contains 192 systolic arrays of size $64\times32$ and D2D PHYs provisioned with 256 GB/s bandwidth. 
Along three sides of the interposer, 12 memory-PHY chiplets interface with 24 off-package GDDR7 stacks, with each memory-PHY chiplet connected to two GDDR7 stacks (8 GB each~\cite{jedec2024jesd239a}).
This configuration provides a total of 192 GB capacity and enough bandwidth for compute-bound prefill execution.

\begin{table}[b]
\vspace{-3mm}
\caption{Comparison of DUET prefill and decode packages against a B200 GPU.}
\label{tab:spec}
\vspace{-4mm}
\resizebox{\linewidth}{!}{%
\begin{tabular}{@{}c|cc|c@{}}
\toprule
 & \textbf{DUET Prefill} & \textbf{DUET Decode} & \textbf{B200~\cite{nvidia_blackwell_datasheet_2024}} \\ \hline
\begin{tabular}{@{}c@{}}\textbf{Computing}\\ \textbf{Unit}\end{tabular} & \begin{tabular}[c]{@{}c@{}}Systolic Array \\ (64$\times$32)\end{tabular} & \begin{tabular}[c]{@{}c@{}}Vector Unit Array \\ (32-wide 16$\times$8)\end{tabular} & \begin{tabular}[c]{@{}c@{}}Tensor Core \\ (8$\times$8$\times$16)\end{tabular} \\
\textbf{Total units} &\hspace{-1mm}192 array $\times$ 16 chiplets & 96 array $\times$ 8 chiplets\hspace{-1mm}& 640 \\
\textbf{Frequency} & 700MHz & 700MHz & 1.8GHz \\
\textbf{FP16 Peak Perf.} & 4.4 PFLOPS/s & 2.2 PFLOPS/s & 2.3 PFLOPS/s \\ \hline
\textbf{Memory type} & GDDR7 (128GB/s) & HBM3e (1TB/s) & HBM3e (1TB/s) \\
\textbf{Memory Cap.} & 24x8 = 192 GB & 12x24 = 288GB & 8x24 = 192 GB \\
\textbf{Memory BW} & 3TB/s & 12TB/s & 8TB/s \\ \bottomrule
\end{tabular}
}
\vspace{-6mm}
\end{table}

\bh{Decode package:}
The Decode package integrates 8 vector-unit chiplets paired with 12 in-package HBM3e stacks in a compact $4\times5$ arrangement.
Each compute chiplet contains 96 32-wide $16\times8$ vector-unit arrays. 
Three HBM columns interleave with two compute-chiplet columns to shorten routing paths and maximize bandwidth. D2D links operate at 1024 GB/s.

\bh{Baseline Systems:}
We compare DUET to three systems. 
(1) NVIDIA B200~\cite{nvidia_blackwell_datasheet_2024}  GPUs configured with tensor-core–based matrix multiplication (using an assumed $8\times8\times16$ configuration to match the peak FP16 performance), as listed in Table~\ref{tab:spec}. 
This system represents the software-based disaggregated inference baseline~\cite{zhong2024distserve}, where prefill and decode phases are split across two homogeneous GPUs.
(2) A prefill-friendly aggregated system that matches the geometry and memory subsystem of the DUET Prefill package, but equips every compute chiplet with both systolic and vector unit arrays. 
(3) A decode-friendly aggregated system that matches the geometry and memory subsystem of the DUET Decode package, with each compute chiplet similarly integrating both compute types.

\begin{figure*}[t]
\centering
\centering
\includegraphics[width=\linewidth]{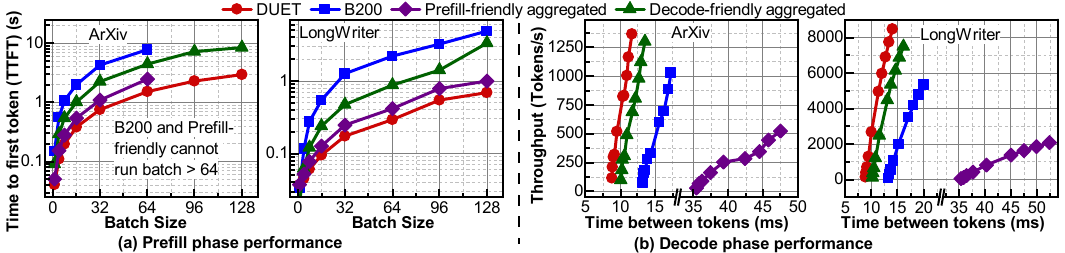}
\vspace{-7mm}
\caption{Performance evaluation of Nemotron-H-56B model on ArXiv and LongWriter workloads.} 
\vspace{-3mm}
\label{fig:llm_perf}
\end{figure*}  

\begin{table*}[t]
\caption{LLM inference performance summary. The Normalized Geo-mean is computed by normalizing each baseline to DUET across all models and workloads, then taking the geometric mean per row.}
\vspace{-3mm}
\label{tab:perf}
\resizebox{\linewidth}{!}{%
\begin{tabular}{l|c|cccc|cccc|cccc|c}
\hline
\multirow{2}{*}{\textbf{Systems}} & \multirow{2}{*}{\textbf{\begin{tabular}[c]{@{}c@{}}Performance \\ Metrics\end{tabular}}} & \multicolumn{4}{c|}{\textbf{Nemotron-H-56B}} & \multicolumn{4}{c|}{\textbf{Zamba2-7B}} & \multicolumn{4}{c|}{\textbf{Llama3-8B}} & \multicolumn{1}{l}{\multirow{2}{*}{\textbf{\begin{tabular}[c]{@{}c@{}}\textbf{Normalized}\\ \textbf{Geo-mean}\end{tabular}}}} \\ \cline{3-14} 
 &  & \multicolumn{1}{c}{\textbf{ArXiv}} & \multicolumn{1}{c}{\textbf{BWB}} & \multicolumn{1}{c}{\textbf{Chat}} & \multicolumn{1}{c|}{\textbf{Longwriter}} & \multicolumn{1}{c}{\textbf{ArXiv}} & \multicolumn{1}{c}{\textbf{BWB}} & \multicolumn{1}{c}{\textbf{Chat}} & \multicolumn{1}{c|}{\textbf{Longwriter}} & \multicolumn{1}{c}{\textbf{ArXiv}} & \multicolumn{1}{c}{\textbf{BWB}} & \multicolumn{1}{c}{\textbf{Chat}} & \multicolumn{1}{c|}{\textbf{Longwriter}}& \multicolumn{1}{l}{} \\ \hline
\textbf{DUET} & \multirow{4}{*}{\textbf{TTFT (ms) $\downarrow$}} & \textbf{2136} & \textbf{3110} & \textbf{102} & \textbf{415} & \textbf{1059} & \textbf{1299} & \textbf{43} & \textbf{169} & \textbf{293} & \textbf{93} & \textbf{12} & \textbf{45} & \textbf{1$\times$} \\
\textbf{B200} &  & 12560 & 18864 & 521 & 2765 & 3015 & 3831 & 104 & 461 & 1304 & 381 & 42 & 195 & \textbf{4.0$\times$} \\
\textbf{Prefill Friendly Agg.} &  & 3332 & 4182 & 137 & 597 & 1380 & 1731 & 54 & 220 & 498 & 155 & 17 & 73 & \textbf{1.4$\times$} \\ 
\textbf{Decode Friendly Agg.} &  & 6140 & 8213 & 247 & 1471 & 1619 & 2317 & 63 & 265 & 948 & 1229 & 30 & 138 & \textbf{2.7$\times$} \\ \hline
\textbf{DUET} & \multirow{4}{*}{\textbf{\begin{tabular}[c]{@{}c@{}}Throughput\\ (Tokens/sec) $\uparrow$\end{tabular}}} & \textbf{621} & \textbf{3340} & \textbf{1359} & \textbf{3431} & \textbf{2283} & \textbf{5197} & \textbf{6368} & \textbf{5412} & \textbf{5002} & \textbf{12946} & \textbf{11905} & \textbf{11608} & \textbf{1$\times$} \\
\textbf{B200} &  & 447 & 2170 & 937 & 2489 & 1490 & 3449 & 4249 & 4142 & 3320 & 8794 & 9573 & 8862 & \textbf{0.7$\times$}\\
\textbf{Prefill Friendly Agg.} &  & 240 & 937 & 399 & 974 & 590 & 1336 & 1614 & 2387 & 1291 & 3500 & 3654 & 4066 & \textbf{0.3$\times$} \\ 
\textbf{Decode Friendly Agg.} &  & 540 & 2926 & 1163 & 3125 & 1975 & 4502 & 5558 & 4935 & 4348 & 11394 & 11311 & 11007 & \textbf{0.9$\times$} \\ \hline
\textbf{DUET} & \multirow{4}{*}{\textbf{\begin{tabular}[c]{@{}c@{}}Time Between \\ Tokens (ms) $\downarrow$\end{tabular}}} & \textbf{9.85} & \textbf{10.98} & \textbf{9.66} & \textbf{10.49} & \textbf{3.34} & \textbf{7.05} & \textbf{2.06} & \textbf{4.08} & \textbf{1.40} & \textbf{2.52} & \textbf{0.87} & \textbf{1.63} & \textbf{1$\times$} \\
\textbf{B200} &  & 14.87 & 16.46 & 14.52 & 15.84 & 5.02 & 10.80 & 3.13 & 6.41 & 2.06 & 3.84 & 1.31 & 2.50 & \textbf{1.5$\times$} \\
\textbf{Prefill Friendly Agg.} &  & 40.31 & 44.07 & 38.77 & 41.95 & 13.44 & 29.14 & 8.35 & 15.48 & 5.20 & 10.55 & 3.51 & 6.52 & \textbf{4.0$\times$} \\ 
\textbf{Decode Friendly Agg.} &  & 11.24 & 12.59 & 11.08 & 12.07 & 3.76 & 8.10 & 2.37 & 4.73 & 1.60 & 2.91 & 1.00 & 1.90 & \textbf{1.2$\times$} \\ \hline
\end{tabular}
}
\vspace{-3mm}
\end{table*}

\vspace{-6mm}
\subsection{Performance Evaluation on LLM Workloads}\label{ssec:perf_eval}
This section evaluates system performance using three metrics: time to first token for the prefill phase; throughput (tokens/s), and average time between tokens for the decode phase, measured across batch sizes from 1 to 128.
Following prior LLM serving frameworks, such as DistServe~\cite{zhong2024distserve} and Mooncake~\cite{qin2025mooncake}, we assume a continuous stream of user requests. 
Each new request is processed on the Prefill package, after which its KV- and state-cache are immediately transferred to the Decode package for token generation.
While we do not explicitly model multi-request scheduling, we assume that the throughput of prefill and decode pipelines is matched, ensuring both packages remain fully utilized.

\bh{Prefill Phase Performance:}
Figure~\ref{fig:llm_perf}(a) plots the TTFT of the Nemotron-H-56B on the ArXiv and LongWriter datasets. 
DUET architecture consistently achieves the lowest TTFT for all batch sizes. 
While running the ArXiv dataset, DUET achieves on average 1.6$\times$ and 2.9$\times$ lower TTFT than prefill- and decode-friendly configurations, respectively. Moreover, DUET has 5.9$\times$ lower TTFT than the B200 GPU baseline because of higher peak performance and SSM/GEMM optimized systolic array. 
The B200 GPU and the prefill-friendly aggregated baselines cannot run with batch sizes over 64 for the ArXiv workload due to KV/state cache capacity limits. 
In contrast, DUET sustains the full batch range because the Prefill package stream caches to the Decode package concurrently, avoiding storage bottlenecks.
Similarly, DUET achieves 1.1-6.7$\times$ lower TTFT than the baselines on the LongWriter dataset.

As summarized in Table~\ref{tab:perf}, DUET outperforms the baselines across all models and datasets. 
Using the geometric mean across all models and workloads, DUET reduces TTFT by 4$\times$, 1.4$\times$, and 2.7$\times$ relative to the B200, prefill-friendly, and decode-friendly systems.

\bh{Decode Phase Performance:}
Figure~\ref{fig:llm_perf}(b) shows the decode throughput–latency trade-off for Nemotron-H-56B on the ArXiv and LongWriter datasets.
Each point in the plot corresponds to a different batch size.
As batch size increases, throughput improves, but the TBT also grows. 
The slope of each curve captures the throughput gained per unit of TBT cost; thus, steeper slopes and points closer to the upper-left region are preferred. 
DUET consistently dominates all baselines across both datasets, maintaining the highest slope and offering the most favorable trade-off.
On the ArXiv dataset, DUET achieves 1.4$\times$, 2.6$\times$, and 1.2$\times$ higher throughput than B200, prefill-friendly, and decode-friendly systems. 
Similarly, DUET achieves 1.1-3.9$\times$ higher throughput than the baselines on the LongWriter dataset.
The decode-friendly aggregated system is the closest competitor due to its similar memory bandwidth; however, it stalls on vector SSM/GEMV operations because half of its compute units are systolic (although we opportunistically utilize systolic arrays as well).
The B200 GPU trails further because decode is predominantly bandwidth-bound. 
The prefill-friendly system, which was competitive during the prefill phase, now falls significantly behind due to its limited memory bandwidth.

As summarized in Table~\ref{tab:perf}, DUET’s advantages extend broadly across all evaluated models and datasets. 
Aggregating results via geometric mean, DUET achieves 1.4$\times$ / 3.4$\times$ / 1.1$\times$ higher throughput over B200, prefill-friendly and decode-friendly systems, respectively, while reducing TBT by 1.5$\times$ / 4.0$\times$ / 1.1$\times$.

\vspace{-4mm}
\subsection{Power, Frequency and Area  Analysis}\label{ssec:area_power}


This section evaluates the power and frequency of the proposed microarchitectures using synthesized systolic and vector-unit arrays.
We then assess system feasibility by estimating the compute-chiplet area at 7 nm to ensure competitiveness with modern semiconductor nodes.
To establish baseline power, area, and timing, we synthesize a 64$\times$32 systolic array and a 32-wide 16$\times$8 vector-unit array in TSMC 28 nm.
As shown in Table~\ref{tab:area_power}, both arrays reach a maximum operating frequency of 763 MHz. 
We conservatively adopt a lower operating frequency of 700 MHz (Table~\ref{tab:spec}) to account for physical-design margin.
The systolic array occupies 3.01 $mm^2$ and consumes 935 mW, whereas the vector-unit array occupies 6.61 $mm^2$ and consumes 3446 mW.

To estimate complete chiplet footprints, SRAM buffer areas are derived from CACTI~\cite{balasubramonian2017cacti} (32 nm): 4.51 $mm^2$ for the systolic-array (1 MB, 256 GB/s) and 6.55 $mm^2$ for the vector-unit array (1 MB, 1024 GB/s).
SFUs are modeled using FlexSFU~\cite{reggiani2023flex} (28 nm), adding 0.95 $mm^2$ for 64 units in the systolic and 3.8 $mm^2$ for 256 units in the vector-unit array.
Die-to-die links follow Nvidia GRS~\cite{poulton20181}: each 8-lane, 25-Gbps PHY occupies 0.3876 $mm^2$ in 16 nm, with the Prefill and Decode chiplets requiring 11 and 42 PHYs, respectively.
Micro-bump area follows UCIe x64 advanced-package assumptions~\cite{sharma2022universal}.
We scale all components to 7 nm using DeepScale~\cite{sarangi2021deepscaletool} and reserve 40\% additional area for controllers, routing, and physical-design overhead.
Under these assumptions, as reported in Table~\ref{tab:spec}, a 121-$mm^2$ Prefill chiplet can integrate up to 192 systolic arrays, and a same-sized Decode chiplet can integrate 96 vector-unit arrays.



\begin{table}[t]
\caption{Power, frequency, and area characteristics of systolic and vector-unit arrays synthesized in TSMC 28 nm.}
\vspace{-4mm}
\label{tab:area_power}
\centering
\resizebox{\linewidth}{!}{%
\begin{tabular}{@{}c|cc@{}}
\toprule
\textbf{Array type} & \textbf{Systolic Array (64$\times$32)} & \textbf{Vector Unit Array (32$\times$16$\times$8)} \\ \hline
\textbf{Power (mW)} & 935 & 3446 \\ 
\textbf{Frequency (MHz)} & 763 & 763 \\
\textbf{Area (mm$^2$)} & 3.01 & 6.61 \\ \bottomrule
\end{tabular}%
}
\vspace{-5mm}
\end{table}

\vspace{-2mm}
\section{Conclusion} \label{sec:conclusion}
DUET introduces a disaggregated hybrid Mamba–Transformer acceleration framework that aligns hardware resources with the inherent compute–memory asymmetry of LLM inference.
By pairing compute-optimized systolic arrays for the prefill phase with bandwidth-optimized vector-unit arrays for the decode phase, DUET efficiently supports both attention and SSM kernels through unified, runtime-configurable microarchitectures.
This co-design across system and circuit levels enables higher performance in both phases while preserving flexibility for emerging hybrid models.
Evaluations demonstrate consistent improvements in TTFT, throughput, and TBT over strong GPU and aggregated-systems baselines, confirming the practicality and scalability of phase-specialized disaggregation.
DUET provides both a concrete design point and a blueprint for future large-scale hybrid LLM acceleration.

\vspace{2pt}
\noindent \textit{Disclosure}: Dr. Ogras serves as a contractor for Samsung Austin Research \& Development Center and Advanced Computing Lab (SARC/ACL). This relationship has been approved under applicable outside activities policies.



\bibliographystyle{ACM-Reference-Format}
\bibliography{ref/sample-base}

@inproceedings{zhong2024distserve,
  title={$\{$DistServe$\}$: Disaggregating prefill and decoding for goodput-optimized large language model serving},
  author={Zhong, Yinmin and Liu, Shengyu and Chen, Junda and Hu, Jianbo and Zhu, Yibo and Liu, Xuanzhe and Jin, Xin and Zhang, Hao},
  booktitle={18th USENIX Symposium on Operating Systems Design and Implementation (OSDI 24)},
  pages={193--210},
  year={2024}
}

@inproceedings{qin2025mooncake,
  title={Mooncake: Trading more storage for less computation—a $\{$KVCache-centric$\}$ architecture for serving $\{$LLM$\}$ chatbot},
  author={Qin, Ruoyu and Li, Zheming and He, Weiran and Cui, Jialei and Ren, Feng and Zhang, Mingxing and Wu, Yongwei and Zheng, Weimin and Xu, Xinran},
  booktitle={23rd USENIX Conference on File and Storage Technologies (FAST 25)},
  pages={155--170},
  year={2025}
}

@article{blakeman2025nemotron,
  title={Nemotron-H: A Family of Accurate and Efficient Hybrid Mamba-Transformer Models},
  author={Blakeman, Aaron and others},
  journal={arXiv preprint arXiv:2504.03624},
  year={2025}
}

@article{lieber2024jamba,
  title={Jamba: A hybrid transformer-mamba language model},
  author={Lieber, Opher and others},
  journal={arXiv preprint arXiv:2403.19887},
  year={2024}
}

@article{glorioso2024zamba,
  title={Zamba: A compact 7b ssm hybrid model},
  author={Glorioso, Paolo and Anthony, Quentin and Tokpanov, Yury and Whittington, James and Pilault, Jonathan and Ibrahim, Adam and Millidge, Beren},
  journal={arXiv preprint arXiv:2405.16712},
  year={2024}
}

@article{glorioso2024zamba2,
  title={The zamba2 suite: Technical report},
  author={Glorioso, Paolo and Anthony, Quentin and Tokpanov, Yury and Golubeva, Anna and Shyam, Vasudev and Whittington, James and Pilault, Jonathan and Millidge, Beren},
  journal={arXiv preprint arXiv:2411.15242},
  year={2024}
}

@misc{ibm_bamba_2025,
  title        = {IBM crossed a transformer with an SSM and got “Bamba”},
  author       = {Martineau, Kim},
  year         = {2025},
  month        = apr,
  publisher    = {IBM Research},
  howpublished = {\url{https://research.ibm.com/blog/bamba-ssm-transformer-model}}
}

@article{grattafiori2024llama3,
  author       = {Grattafiori, Alessandro and others},
  title        = {The {\textsc Llama 3} Herd of Models},
  journal      = {arXiv preprint},
  volume       = {arXiv:2407.21783},
  year         = {2024},
  url          = {https://arxiv.org/abs/2407.21783}
}

@inproceedings{gu2024mamba,
  title={Mamba: Linear-time sequence modeling with selective state spaces},
  author={Gu, Albert and Dao, Tri},
  booktitle={First conference on language modeling},
  year={2024}
}

@article{dao2024transformers,
  title={Transformers are ssms: Generalized models and efficient algorithms through structured state space duality},
  author={Dao, Tri and Gu, Albert},
  journal={arXiv preprint arXiv:2405.21060},
  year={2024}
}

@misc{amd_mi350_series_2025,
  title        = {AMD Instinct MI350 Series GPUs: A Game Changer for Inference, Training and HPC Workloads},
  author       = {AMD Corporation},
  year         = {2025},
  month        = sep,
  howpublished = {\url{https://www.amd.com/en/blogs/2025/amd-instinct-mi350-series-game-changer.html}},
  note         = {Accessed: 2025-11-13}
}

@misc{nvidia_blackwell_datasheet_2024,
  title        = {NVIDIA Blackwell Architecture},
  author       = {NVIDIA Corporation},
  year         = {2024},
  howpublished = {\url{https://resources.nvidia.com/en-us-blackwell-architecture/datasheet}},
  note         = {Accessed: 2025-11-13}
}

@article{zhang2025spad,
  title={SPAD: Specialized Prefill and Decode Hardware for Disaggregated LLM Inference},
  author={Zhang, Hengrui and Patel, Pratyush and Ning, August and Wentzlaff, David},
  journal={arXiv preprint arXiv:2510.08544},
  year={2025}
}

@inproceedings{li2024marca,
  title={Marca: Mamba accelerator with reconfigurable architecture},
  author={Li, Jinhao and Huang, Shan and Xu, Jiaming and Liu, Jun and Ding, Li and Xu, Ningyi and Dai, Guohao},
  booktitle={Proceedings of the 43rd IEEE/ACM International Conference on Computer-Aided Design},
  pages={1--9},
  year={2024}
}

@article{li2025marca,
  title={MARCA-v2: Mamba Accelerator with Complementary State Space Model Sparsity and Reconfigurable Architecture},
  author={Li, Jinhao and Huang, Shan and Xu, Jiaming and Liu, Jun and Xu, Ningyi and Dai, Guohao},
  journal={IEEE Transactions on Computer-Aided Design of Integrated Circuits and Systems},
  year={2025},
  publisher={IEEE}
}

@inproceedings{jin2025hcsas,
  title={HCSAs: Hybrid Computing Systolic Arrays for Accelerating Mamba Models with Unified State Space Buffers and Energy-Efficient Dataflow},
  author={Jin, Xu and Zheng, Haotian and Nie, Maohua and Wang, Jialin and Shi, C-J Richard},
  booktitle={Proceedings of the Great Lakes Symposium on VLSI 2025},
  pages={457--462},
  year={2025}
}

@inproceedings{jung2025hlx,
  title={HLX: A Unified Pipelined Architecture for Optimized Performance of Hybrid Transformer-Mamba Language Models},
  author={Jung, In-Jun and Yang, Gyeongrok and Min, Jaeha and Kim, Joo-Young},
  booktitle={Proceedings of the 58th IEEE/ACM International Symposium on Microarchitecture{\textregistered}},
  pages={461--475},
  year={2025}
}

@article{raja2025systolic,
  title={Systolic Array-based Accelerator for Structured State-Space Models},
  author={Raja, Shiva and Demirkiran, Cansu and Sarkar, Aakash and Popovic, Milos and Joshi, Ajay},
  journal={arXiv preprint arXiv:2507.21394},
  year={2025}
}

@inproceedings{wei2025lightmamba,
  title={Lightmamba: Efficient mamba acceleration on fpga with quantization and hardware co-design},
  author={Wei, Renjie and Xu, Songqiang and Zhong, Linfeng and Yang, Zebin and Guo, Qingyu and Wang, Yuan and Wang, Runsheng and Li, Meng},
  booktitle={2025 Design, Automation \& Test in Europe Conference (DATE)},
  pages={1--7},
  year={2025},
  organization={IEEE}
}

@article{wang2025fastmamba,
  title={FastMamba: A High-Speed and Efficient Mamba Accelerator on FPGA with Accurate Quantization},
  author={Wang, Aotao and Shao, Haikuo and Ma, Shaobo and Wang, Zhongfeng},
  journal={arXiv preprint arXiv:2505.18975},
  year={2025}
}

@article{ko2025ssm,
  title={SSM-RDU: A Reconfigurable Dataflow Unit for Long-Sequence State-Space Models},
  author={Ko, Sho and Olukotun, Kunle},
  journal={arXiv preprint arXiv:2503.22937},
  year={2025}
}

@article{kim2025emamba,
  title={eMamba: Efficient Acceleration Framework for Mamba Models in Edge Computing},
  author={Kim, Jiyong and others},
  journal={ACM Transactions on Embedded Computing Systems},
  volume={24},
  number={5s},
  pages={1--22},
  year={2025},
  publisher={ACM New York, NY}
}

@article{zhong2025specmamba,
  title={SpecMamba: Accelerating Mamba Inference on FPGA with Speculative Decoding},
  author={Zhong, Linfeng and Xu, Songqiang and Wen, Huifeng and Xie, Tong and Guo, Qingyu and Wang, Yuan and Li, Meng},
  journal={arXiv preprint arXiv:2509.19873},
  year={2025}
}

@inproceedings{stow2016cost,
  title={Cost analysis and cost-driven IP reuse methodology for SoC design based on 2.5 D/3D integration},
  author={Stow, Dylan and Akgun, Itir and Barnes, Russell and Gu, Peng and Xie, Yuan},
  booktitle={2016 IEEE/ACM International Conference on Computer-Aided Design (ICCAD)},
  pages={1--6},
  year={2016},
  organization={IEEE}
}

@article{sharma2025hemu,
  title={HeMu: Energy-Efficient DNN Inferencing via Heterogeneous-Multi-Chiplet Architectures},
  author={Sharma, Harsh and Kanani, Alish and Doppa, Janardhan Rao and Ogras, Umit Y and Pande, Partha Pratim},
  journal={IEEE Transactions on Computer-Aided Design of Integrated Circuits and Systems},
  year={2025},
  publisher={IEEE}
}

@article{kanani2025thermos,
  title={THERMOS: Thermally-aware multi-objective scheduling of AI workloads on heterogeneous multi-chiplet PIM architectures},
  author={Kanani, Alish and Pfromm, Lukas and Sharma, Harsh and Doppa, Jana and Pande, Partha and Ogras, Umit},
  journal={ACM Transactions on Embedded Computing Systems},
  volume={24},
  number={5s},
  pages={1--26},
  year={2025},
  publisher={ACM New York, NY}
}

@article{taheri2024red,
  title={Red: A reliable and deadlock-free routing for 2.5-d chiplet-based interposer networks},
  author={Taheri, Ebadollah and Pasricha, Sudeep and Nikdast, Mahdi},
  journal={IEEE Transactions on Computer-Aided Design of Integrated Circuits and Systems},
  volume={43},
  number={12},
  pages={4599--4612},
  year={2024},
  publisher={IEEE}
}

@inproceedings{qi2023moela,
  title={Moela: A multi-objective evolutionary/learning design space exploration framework for 3d heterogeneous manycore platforms},
  author={Qi, Sirui and Li, Yingheng and Pasricha, Sudeep and Kim, Ryan Gary},
  booktitle={2023 Design, Automation \& Test in Europe Conference \& Exhibition (DATE)},
  pages={1--6},
  year={2023},
  organization={IEEE}
}

@article{jaiswal2024halo,
  title={HALO: Communication-aware Heterogeneous 2.5 D System for Energy-efficient LLM Execution at Edge},
  author={Jaiswal, Abhi and Shahana, KC Sharin and Ravichandran, Sujitha and Adarsh, K and Bhat, H Bharath and Joardar, Biresh Kumar and Mandal, Sumit K},
  journal={IEEE Journal on Emerging and Selected Topics in Circuits and Systems},
  year={2024},
  publisher={IEEE}
}

@inproceedings{ayari2016schedulability,
  title={Schedulability-guided exploration of multi-core systems},
  author={Ayari, Rabeh and Hafnaoui, Imane and Beltrame, Giovanni and Nicolescu, Gabriela},
  booktitle={Proceedings of the 27th International Symposium on Rapid System Prototyping: Shortening the Path from Specification to Prototype},
  pages={121--127},
  year={2016}
}

@inproceedings{wei20239,
  title={9.3 NVLink-C2C: A coherent off package chip-to-chip interconnect with 40Gbps/pin single-ended signaling},
  author={Wei, Ying and Huang, Yi Chieh and Tang, Haiming and Sankaran, Nithya and Chadha, Ish and Dai, Dai and Oluwole, Olakanmi and Balan, Vishnu and Lee, Edward},
  booktitle={2023 IEEE International Solid-State Circuits Conference (ISSCC)},
  pages={160--162},
  year={2023},
  organization={IEEE}
}

@misc{ualink2025_200g,
  title        = {Introducing UALink 200G 1.0 Specification},
  author       = {Ultra Accelerator Link Consortium},
  year         = {2025},
  month        = apr,
  howpublished = {\url{https://ualinkconsortium.org/wp-content/uploads/2025/04/UALink-1.0-White_Paper_FINAL.pdf}},
  note         = {Accessed: 2025-11-13}
}

@inproceedings{kung1979systolic,
  title={Systolic arrays (for VLSI)},
  author={Kung, Hsiang Tsung and Leiserson, Charles E},
  booktitle={Sparse Matrix Proceedings 1978},
  volume={1},
  pages={256--282},
  year={1979},
  organization={Society for industrial and applied mathematics Philadelphia, PA, USA}
}

@inproceedings{genc2021gemmini,
  title={Gemmini: Enabling systematic deep-learning architecture evaluation via full-stack integration},
  author={Genc, Hasan and others},
  booktitle={2021 58th ACM/IEEE Design Automation Conference (DAC)},
  pages={769--774},
  year={2021},
  organization={IEEE}
}

@article{park2022192,
  title={A 192-Gb 12-high 896-GB/s HBM3 DRAM with a TSV auto-calibration scheme and machine-learning-based layout optimization},
  author={Park, Myeong-Jae and others},
  journal={IEEE Journal of Solid-State Circuits},
  volume={58},
  number={1},
  pages={256--269},
  year={2022},
  publisher={IEEE}
}

@techreport{jedec2024jesd239a,
  title        = {JESD239A: Graphics Double Data Rate (GDDR7) SGRAM Devices},
  institution  = {JEDEC Solid State Technology Association},
  year         = {2024},
  note         = {JEDEC standard for GDDR7 graphics memory devices},
  howpublished = {\url{https://store.accuristech.com/standards/jedec-jesd239a}}
}

@article{cohan2018discourse,
  title={A discourse-aware attention model for abstractive summarization of long documents},
  author={Cohan, Arman and others},
  journal={arXiv preprint arXiv:1804.05685},
  year={2018}
}

@article{jiang2023discourse,
  title={Discourse centric evaluation of machine translation with a densely annotated parallel corpus},
  author={Jiang, Yuchen Eleanor and Liu, Tianyu and Ma, Shuming and Zhang, Dongdong and Sachan, Mrinmaya and Cotterell, Ryan},
  journal={arXiv preprint arXiv:2305.11142},
  year={2023}
}

@inproceedings{bailongwriter,
  title={LongWriter: Unleashing 10,000+ Word Generation from Long Context LLMs},
  author={Bai, Yushi and others},
  booktitle={The Thirteenth International Conference on Learning Representations}
}

@article{zheng2023lmsys,
  title={Lmsys-chat-1m: A large-scale real-world llm conversation dataset},
  author={Zheng, Lianmin and others},
  journal={arXiv preprint arXiv:2309.11998},
  year={2023}
}

@article{lau2022recent,
  title={Recent advances and trends in advanced packaging},
  author={Lau, John H},
  journal={IEEE Transactions on Components, Packaging and Manufacturing Technology},
  volume={12},
  number={2},
  pages={228--252},
  year={2022},
  publisher={IEEE}
}

@article{balasubramonian2017cacti,
  title={CACTI 7: New tools for interconnect exploration in innovative off-chip memories},
  author={Balasubramonian, Rajeev and Kahng, Andrew B and Muralimanohar, Naveen and Shafiee, Ali and Srinivas, Vaishnav},
  journal={ACM Transactions on Architecture and Code Optimization (TACO)},
  volume={14},
  number={2},
  pages={1--25},
  year={2017},
  publisher={ACM New York, NY, USA}
}

@article{luo2023ramulator,
  title={Ramulator 2.0: A modern, modular, and extensible dram simulator},
  author={Luo, Haocong and Tu{\u{g}}rul, Yahya Can and Bostanc{\i}, F Nisa and Olgun, Ataberk and Ya{\u{g}}l{\i}k{\c{c}}{\i}, A Giray and Mutlu, Onur},
  journal={IEEE Computer Architecture Letters},
  volume={23},
  number={1},
  pages={112--116},
  year={2023},
  publisher={IEEE}
}

@inproceedings{reggiani2023flex,
  title={Flex-sfu: Accelerating dnn activation functions by non-uniform piecewise approximation},
  author={Reggiani, Enrico and Andri, Renzo and Cavigelli, Lukas},
  booktitle={2023 60th ACM/IEEE Design Automation Conference (DAC)},
  pages={1--6},
  year={2023},
  organization={IEEE}
}

@article{poulton20181,
  title={A 1.17-pJ/b, 25-Gb/s/pin ground-referenced single-ended serial link for off-and on-package communication using a process-and temperature-adaptive voltage regulator},
  author={Poulton, John W and others},
  journal={IEEE Journal of Solid-State Circuits},
  volume={54},
  number={1},
  pages={43--54},
  year={2018},
  publisher={IEEE}
}

@article{sharma2022universal,
  title={Universal chiplet interconnect express (UCIe): An open industry standard for innovations with chiplets at package level},
  author={Sharma, Debendra Das and Pasdast, Gerald and Qian, Zhiguo and Aygun, Kemal},
  journal={IEEE Transactions on Components, Packaging and Manufacturing Technology},
  volume={12},
  number={9},
  pages={1423--1431},
  year={2022},
  publisher={IEEE}
}

@inproceedings{sarangi2021deepscaletool,
  title={DeepScaleTool: A tool for the accurate estimation of technology scaling in the deep-submicron era},
  author={Sarangi, Satyabrata and Baas, Bevan},
  booktitle={2021 IEEE International Symposium on Circuits and Systems (ISCAS)},
  pages={1--5},
  year={2021},
  organization={IEEE}
}


\end{document}